\documentclass[12pt]{spieman}  %
\usepackage{amsmath,amsfonts,amssymb}
\usepackage{graphicx}
\usepackage{setspace}
\usepackage{tocloft}

\usepackage{bm}

\definecolor{purple}{rgb}{0.5,0.0,1.0}
\newcommand{\ch}[1]{#1}  %

\newcommand{\bi}{\begin{itemize}}
\newcommand{\ei}{\end{itemize}}
\newcommand{\ben}{\begin{enumerate}}
\newcommand{\een}{\end{enumerate}}
\newcommand{\be}{\begin{equation}}
\newcommand{\ee}{\end{equation}}
\newcommand{\bea}{\begin{eqnarray}} 
\newcommand{\eea}{\end{eqnarray}}
\newcommand{\ba}{\begin{align}} 
\newcommand{\ea}{\end{align}}
\newcommand{\bse}{\begin{subequations}} 
\newcommand{\ese}{\end{subequations}}
\newcommand{\bc}{\begin{center}}
\newcommand{\ec}{\end{center}}
\newcommand{\bfi}{\begin{figure}}
\newcommand{\efi}{\end{figure}}
\newcommand{\ca}[2]{\caption{#1 \label{#2}}}
\newcommand{\ig}[2]{\includegraphics[#1]{#2}}
\newcommand{\bmp}[1]{\begin{minipage}{#1}}
\newcommand{\emp}{\end{minipage}}

\newcommand{\tbox}[1]{{\mbox{\scriptsize #1}}}
\newcommand{\mbf}[1]{{\mathbf #1}}

\newcommand{\RR}{\mathbb{R}}
\newcommand{\ZZ}{\mathbb{Z}}
\newcommand{\eps}{\varepsilon}
\newcommand{\bigO}{{\mathcal O}}

\newcommand{\etal}{et al.\ }
\newcommand{\emach}{\epsilon_\tbox{mach}}

\newtheorem{thm}{Theorem}

\newtheorem{rmk}[thm]{Remark}
\newcommand{\pO}{{\partial\Omega}}
\newcommand{\xx}{\mbf{x}}
\newcommand{\sss}{\mbf{s}}
\newcommand{\ff}{\mbf{f}}

\newcommand{\rr}{\mbf{r}}
\newcommand{\FF}{\mbf{F}}
\newcommand{\uap}{u^\tbox{ap}}
\newcommand{\uoc}{u^\tbox{oc}}
\newcommand{\uapg}{u^\tbox{ap}_\tbox{geom}}

\newcommand{\fr}{{\mathfrak f}}                   %
\newcommand{\al}{\alpha}

\author[a,*]{Alex H. Barnett}
\affil[a]{Center for Computational Mathematics, Flatiron Institute, Simons Foundation, New York, NY, USA, 10010}

\title{Efficient high-order accurate Fresnel diffraction
  via areal quadrature and the nonuniform FFT}

\cftpagenumbersoff{figure}
\cftpagenumbersoff{table}

\begin{document} 
\maketitle

\begin{abstract}
  We present a fast algorithm for
  computing the diffracted field from arbitrary binary
  (\ch{sharp}-edged) planar apertures
  and occulters in the scalar Fresnel approximation, for up to moderately high
  Fresnel numbers ($\lesssim 10^3$).
  It uses a high-order {\em areal quadrature} over the aperture,
  then exploits a single 2D {\em nonuniform} fast Fourier transform (NUFFT)
  to evaluate rapidly at target points (of order $10^7$ such points per second, independent of aperture complexity).
  It thus combines the high accuracy of edge integral methods
  with the high speed
  of Fourier methods.
  Its cost is $\bigO(n^2 \log n)$,
  where $n$ is the linear resolution required in source and target planes,
  to be compared with $\bigO(n^3)$ for edge integral methods.
  In tests with several aperture shapes this translates to between 2 and 5 orders of magnitude acceleration.
  In starshade modeling for exoplanet astronomy, we find that it is roughly $10^4 \times$ faster than the
  state of the art
  in accurately computing the set of telescope pupil wavefronts.
  We provide a documented, tested MATLAB/Octave implementation.
  
  An appendix shows the mathematical
  equivalence of the boundary diffraction wave,
  angular integration, %
  and line integral %
  formulae, then analyzes a %
  new
  {\em non-singular} reformulation
  that eliminates their common difficulties near the geometric shadow edge.
  This supplies a robust edge integral
  reference against which to validate the main proposal.
\end{abstract}

\keywords{Fresnel, diffraction, quadrature, numerical, nonuniform FFT, starshade}

{\noindent \footnotesize\textbf{*}\linkable{abarnett@flatironinstitute.org} }

\section{Introduction}

The numerical modeling of wave diffraction from thin
two-dimensional (2D) screens and apertures in the Fresnel regime has
many applications in optics
\cite{bornwolf} and acoustics \cite[\S 8.4]{morseingard},
including instrument modeling \cite{poppy,hu20bluestein},
lithography mask design \cite{bourdillon00},
Fourier optics \cite{goodman96},
coherent X-rays \cite{ruizlopez17},
acoustic emission \cite{mast07},
computer-generated binary holograms \cite{tsang11},
starshades \cite{vanderbei07},
and Fresnel zone plate imagers \cite{serre11,wilhem18}.
This usually involves a plane or spherical wavefront
hitting a binary (``0-1'') mask of given shape, although
continuous opacity/phase variation is also possible.
We will confine ourselves to the former case,
although the method can trivially accommodate the latter.
Our point is to present a simple---yet seemingly overlooked---method
which renders their high-accuracy numerical modeling
at moderate Fresnel numbers
orders of magnitude more efficient than before.

One motivation is starshade design
\cite{vanderbei07,lo07,cash11,cady12,shaklan17,harness18}.
Given a space telescope,
the goal is that a distant binary occulter blocks the direct light from a star,
allowing
much dimmer exoplanets separated from it by only tens of milli-arcseconds
to be imaged.
The occulter shape and distance are thus optimized
to give a deep shadow region, with relative
intensity of order $10^{-10}$ across the telescope pupil,
throughout a given wavelength range, while minimizing the occulter's physical size
(for practical reasons), and angular size at the telescope.
This has led to shapes with ``petals'' that emulate
a continuous radial apodization,
radii of order $10$ m, and distances of order $10^7$ m.
Here the scalar \cite{nonscalar} and Fresnel approximations are superb
\cite[App.~A]{harness18},
with a small Fresnel number (defined in \eqref{fn}) of typically $5$--$20$.
Numerical modeling is challenging,
demanding at least 6-digit accuracy in amplitude to validate the shadow,
and many runs with different wavelengths and shapes to
assess mechanical and thermal stability \cite{shaklan17}.

\bfi %
\centering\ig{width=0.4\textwidth}{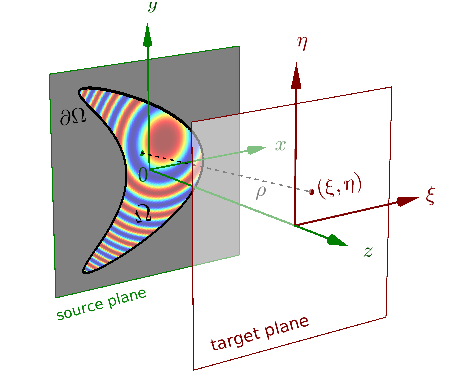}
\vspace{-1ex}
\ca{Geometry and notation for Fresnel diffraction, in the case of $\Omega$
  an aperture with boundary $\pO$.
  A plane wave is incident from behind, along the $z$ axis.
  For the target $(\xi,\eta)$,
  the real part of the integrand in \eqref{fres} is imaged in color
  (red positive,
  blue negative, and green around zero; each red or blue annular
  region is a {\em Fresnel zone}).
}{f:geom}
\efi

Fixing a wavelength $\lambda$ and propagation distance $z$,
a point in the aperture (or occulter) plane is $(x,y,0)$, while a target point
in the detector (or pupil) plane is $(\xi,\eta,z)$; see Fig.~\ref{f:geom}.
We drop the constant $z$-coordinates from now on: the problem
is in essence 2D.
In the case of a unit amplitude plane wave with wavevector $(0,0,2\pi/\lambda)$
incident on a planar aperture $\Omega \subset \RR^2$,
the mathematical task is to evaluate the Fresnel integral
for the scalar potential
\be
\uap(\xi,\eta) \;=\; \frac{1}{i\lambda z} \iint_\Omega e^{\frac{i\pi}{\lambda z}
  \left[ (\xi-x)^2 + (\eta-y)^2 \right] } \,dxdy
~.
\label{fres}
\ee
This takes the form of a 2D convolution of the aperture's
characteristic function $\chi_\Omega$ with a radially-symmetric
kernel (a complex Gaussian), whose half-wave oscillation regions
are commonly called ``zones'' (Fig.~\ref{f:geom}).
If $R$ is an effective (or maximum) radius of $\Omega$,
then the in-plane separation $r := \sqrt{(\xi-x)^2 + (\eta-y)^2}$
is typically bounded by $R$ times a small constant.
Thus the number of zones inside $\Omega$ is of order
\be
\fr \; := \; \frac{R^2}{\lambda z} \qquad \mbox{(Fresnel number)}
~,
\label{fn}
\ee
and the finest oscillation scale of the integrand is $\bigO(1/\fr)$.
In \eqref{fres} the prefactor $1/i\lambda z$ insures that
$u$ tends to unity in the limit of large aperture, ie,
the unimpeded wave.

It is worth
reviewing the origin of
\eqref{fres}.
It arises from the Kirchhoff diffraction approximation
to the full Maxwell equations;
this is good when aperture features are much larger than $\lambda$
\cite[Ch.~3]{goodman96}.
Only the zeroth and first term in the Taylor expansion of
the exponent in the free-space Green's function $e^{2\pi i \rho/\lambda}/\rho$
are then kept, $\rho = \sqrt{r^2 + z^2}$ being
the source-target distance (dotted line in Fig.~\ref{f:geom}).
The Fresnel approximation is thus valid to the extent that the next
term is small, implying the condition $r^4 \ll \lambda z^3$.
The denominator of the Green's function is approximated by $z$.
We refer the reader to \cite[\S8.3.3]{bornwolf} \cite[Ch.~4]{goodman96}
for details.
Note that the zeroth term gave the plane propagation phase
$e^{2\pi i z/\lambda}$, which is usually included as a prefactor in
\eqref{fres}. We drop it for simplicity; it is trivial to insert.
By replacing $z^{-1}$ by $z^{-1} + D^{-1}$,
\eqref{fres} also applies when
a \ch{point source at finite} distance $D$ produces a spherical incident wave 
\cite{bornwolf,dauger96,harness18}.
In the plane wave case, by Babinet's principle
\cite[\S8.3.2]{bornwolf} \cite[\S2.2]{cash11},
the potential when $\Omega$ defines an occulter
rather than an aperture is simply given by
\be
\uoc(\xi,\eta) \;=\; 1 - \uap(\xi,\eta)~.
\label{bab}
\ee

\bfi %
\centering
\ig{width=0.8\textwidth}{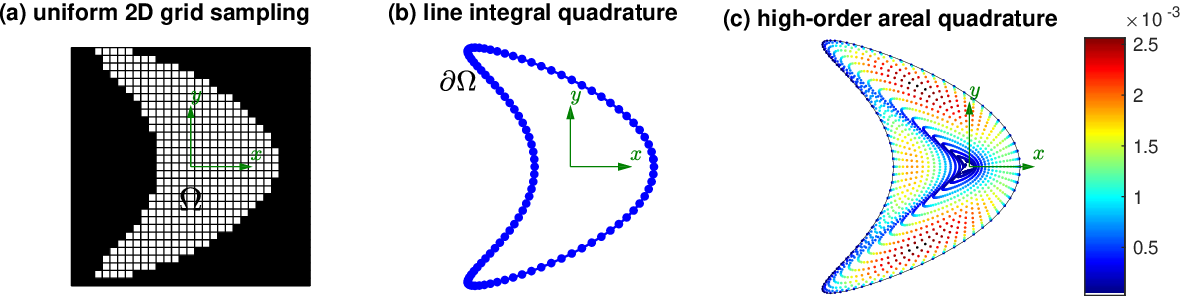}
\vspace{2ex}
\ca{Sketch of three alternative methods for discretization in
  the source (aperture) plane, shown for a
  kite-shaped aperture $\Omega$ with smooth boundary $\pO$.
  (a) and (b) are well established.
  Our method is (c): areal nodes $(x_j,y_j)$ are shown with their color
  indicating the weights $w_j$ according to the scale at the right.
}{f:meth}
\efi

Analytical forms for the integral \eqref{fres}
are known only for the straight edge, infinite slit,
rectangles (all involving the same 1D special function)
\cite[\S8.7]{bornwolf} \cite[\S4.5.1]{goodman96},
\ch{and the disc} \cite{harvey84,sommargren90,dubra99}.
Semi-analytical Bessel expansions can be useful for
symmetric starshade design \cite{vanderbei07,cash11}.
But for general shapes one is left with fully numerical methods,
which
fall into two main categories (sketched in Fig.~\ref{f:meth}(a,b)):
\ben
\item[(a)] 2D Fast Fourier transform (FFT) methods.
  There are two flavors:
  (i) methods implementing the convolution
  theorem (forward then backward FFTs), useful only for
  large $\fr$ (near-field), or (ii)
  methods exploiting the quadratic form in \eqref{fres} via a single FFT
  \cite{lo07,junchang09,serre11},
  useful from zero to moderate $\fr$.
  See \cite{mas99} for a review, where fractional
  FFTs are also considered.
  The FFT of course requires only $\bigO(n^2 \log n)$ operations
  to transform a $n\times n$ source to target grid.
  However, the aperture or occulter must be sampled (quantized) on the source
  grid, and if this is done in a binary fashion \cite{serre11}
  (as in Fig.~\ref{f:meth}(a)),
  the error convergence rate is very low order, no faster than $\bigO(1/n)$.
  This is inadequate for starshade shadow modeling \cite{cash11,cady12}.
  Sub-pixel averaging can improve accuracy \cite{poppy,harness18},
  but this can give at best \cite{meep} $\bigO(1/n^2)$, and
  for starshades huge ($n>10^5$) sized FFTs are still needed
  to reach the needed accuracy \cite{harness18}.
  The underlying problem is that $\chi_\Omega$ is
  {\em not a bandlimited function},
  so is always poorly represented on regular grids.

\item[(b)] Edge integral methods.
  Such methods discretize a target-dependent integral over the
  aperture/occulter boundary $\pO$.
  The literature splits into two formulae:
  (i)
  the ``boundary diffraction wave'' (BDW) of
  Miyamoto--Wolf \cite{miyamoto62},
  arising from the Kirchhoff approximation,
  vs (ii)
  reduction of \eqref{fres} to a 1D angular integral,
  due to Dauger \cite{dauger96} and Dubra--Ferrari \cite{dubra99}.
  To resolve the Fresnel
  integrand, the number of discretization nodes must scale as $n=\bigO(\fr)$,
  so the cost for (direct) evaluation on a resolved $n\times n$ target grid
  is $\bigO(n^3)=\bigO(\fr^3)$.
  Both formulae are applied to
  state of the art starshade modeling:
  Cady \cite{cady12,SISTER} uses (i), while Cash \cite{cash11}
  and Harness \etal \cite{harness18} use (ii).
  Here 2$^\tbox{nd}$-order, ie $\bigO(1/n^2)$ accurate, midpoint quadrature rules
  are used, with up to about $10^5$ nodes, to reach the needed 6-digit accuracy.

  \begin{rmk}Such edge integral methods should not be confused with
  the (more involved)
  boundary integral equation (BIE)
  method, which solves the 3D Helmholtz or Maxwell equations
  using surface unknowns \ch{and potential theory} \cite{coltonkress,lintner3d}.
  \end{rmk}
  \een
  
Our proposal combines the best features of the above two categories,
namely the speed of the FFT methods
with the high (and potentially high-order) accuracy of edge integral methods
for binary apertures.
The result is an acceleration over edge integral methods
of between 2 and 5 orders of magnitude.
In a nutshell the idea is, realizing that a 2D regular grid guarantees poor
quadrature for integrals over $\Omega$, to replace it
with a much better (high-order) {\em areal quadrature scheme};
see Fig.~\ref{f:meth}(c).
Then exploiting the quadratic form in \eqref{fres}, as in
``single FFT'' methods (a)(ii), leaves one remaining problem:
how to rapidly evaluate Fourier sums involving off-grid frequencies.
Fortunately, fast algorithms for this task---nonuniform FFTs\cite{dutt,nufft3}---are quite
mature, and have speeds only one order of magnitude below those of plain FFTs.

High accuracy relies on constructing a good areal quadrature for $\Omega$,
which depends on its geometric description.
We show that such quadratures can easily piggyback off boundary
quadratures, or be built independently.
Our proposal is in some way related to diffraction methods
that subsample the FFT \cite{soummer07} or use chirp FFTs \cite{hu20bluestein};
however, it is much simpler and more general than either.

Turning to the structure of this paper,
Sec.~\ref{s:meth} explains the method for arbitrary
targets (Sec.~\ref{s:t3}), then gridded targets (Sec.~\ref{s:t1}),
the latter being somewhat faster.
In Sec.~\ref{s:test} we demonstrate the high accuracy and efficiency of
the method for a smooth occulter (Sec.~\ref{s:kite}),
two symmetric starshades (Sec.~\ref{s:star}),
and an aperture built from 67 million triangles (Sec.~\ref{s:koch}).
We compare %
to the performance of a BDW edge integral code
\cite{cady12,SISTER}.
We draw conclusions, \ch{explain how the method can be applied to perturbed starshades,} and propose extensions in Sec.~\ref{s:conc}.

Finally,
validating the proposed NUFFT method
down to errors of $10^{-12}$ or better demanded a high-accuracy reference
edge integral method, which required new research, to which the Appendix
is devoted.
There we clarify that {\em the BDW edge-integral (b)(i) and angular-integration (b)(ii) methods are equivalent}, and equivalent to
a more convenient line integral due to Cash \cite{cash11}.
However, as we show, existing edge integral methods suffer numerical breakdown as targets approach $\pO$ (the geometric shadow edge),
because they represent the (smooth)
  diffracted field as a {\em sum of two discontinuous terms, one with a singular
  integrand}.
We instead present and analyze a simple, robust {\em non-singular line integral} (NSLI)
that maintains close to machine accuracy for target points near or even on $\pO$, without extra work, yet takes only five lines to code.

\begin{rmk}
We maintain a documented, tested, open-source MATLAB/Octave
implementation of the proposed fast algorithm
(using the FINUFFT library \cite{finufft}),
and the NSLI, on GitHub \cite{fresnaqgit}.
Queries to the author are welcome.
\end{rmk}

\section{The proposed method}
\label{s:meth}

Given a set of targets $(\xi_k,\eta_k)$, $k=1,\dots,M$,
recall that the goal is
to approximate \eqref{fres} efficiently, ie, to evaluate
\be
\uap_k := \uap(\xi_k,\eta_k) = \frac{1}{i\lambda z} \iint_\Omega e^{\frac{i\pi}{\lambda z}
  \left[ (\xi_k-x)^2 + (\eta_k-y)^2 \right] } \,dxdy, \qquad k=1,\dots,M ~.
\label{fresk}
\ee
Suppose that an {\em areal quadrature rule}
for the aperture $\Omega$ has been found,
that is, a set of nodes $(x_j,y_j) \in \RR^2$ and
weights $w_j$, $j=1,\dots,N$,
such that, for all sufficiently smooth functions $f$,
\be
\iint_\Omega f(x,y) \, dxdy  \; \approx \; \sum_{j=1}^N f(x_j,y_j) w_j
\label{aq}
\ee
holds to high accuracy.
Specifically, one seeks a family of rules of increasing $N$,
with a high order of convergence, $p$, meaning that,
for each $C^\infty$-smooth $f$, the error (difference between
left and right hand sides) is $\bigO(N^{-p})$.
This may even hold for all $p>0$, in which case the convergence is
said to be super-algebraic or {\em spectral}.
For instance, such
a quadrature over a rectangle is given by a tensor product
of 1D Gauss--Legendre rules \cite[Ch.~19]{ATAP}.
By passing those nodes through a smooth mapping
of $\RR^2$ to $\RR^2$ (and multiplying the weights by its Jacobean),
such rules for triangles and distorted quadrilaterals are easily made,
the union of which can approximate any domain with piecewise smooth
boundary to high order.
For more node-efficient
quadratures we refer the reader to recent works
on triangles \cite{vioreanu14}, polygons \cite{mousavi09,xiao10},
and domains defined by piecewise rational curves \cite{gunderman20}.
Converting a {\em general} boundary into an areal quadrature
for its interior is a software engineering task beyond the scope of this paper.
We will be content constructing areal
quadratures for three types of domains: the interior of a smooth closed curve,
symmetric starshades, and unions of triangles.

Since any rule \eqref{aq} must resolve the Fresnel integrand oscillations,
for a fixed $\Omega$ its number of nodes must grow like
\be
N \; =\; \bigO(\fr^2)~.
\label{Ngro}
\ee
Now simply applying \eqref{aq} to \eqref{fresk}
gives a high-order accurate direct (slow) summation method,
that, as the results below show, can exceed accuracy requirements
with reasonable numbers of nodes.%
\footnote{Our results show that claims such as
  ``a single point in the shadow plane
  can require a trillion sine calculations at quadruple precision''
  \cite{cash11} are overly pessimistic.}
The cost of this direct sum to $M$ targets would be $\bigO(NM)$;
our goal is now to reduce this to close to $\bigO(N+M)$
via fast Fourier methods.

\subsection{The fast algorithm for diffraction to arbitrary target points}
\label{s:t3}

We apply \eqref{aq} to \eqref{fresk}, then in the second line expand
each quadratic term,%
\footnote{\ch{This trick, common to ``single FFT'' methods \cite{lo07,junchang09,serre11}, is similar to, but essentially the reverse of, that in the Bluestein method \cite{bluestein70}.}}
to give
\bea
\uap_k &\approx& \frac{1}{i\lambda z} \sum_{j=1}^N
e^{\frac{i\pi}{\lambda z}
  \left[ (\xi_k-x_j)^2 + (\eta_k-y_j)^2 \right] } \,w_j
\label{faq}
\\
& = & \frac{1}{i\lambda z} e^{\frac{i\pi}{\lambda z} (\xi_k^2+\eta_k^2)}
  \cdot
  \sum_{j=1}^N e^{\frac{-2\pi i}{\lambda z} (\xi_k x_j + \eta_k y_j)}
  \biggl( e^{\frac{i\pi}{\lambda z}(x_j^2+y_j^2)}\,w_j \biggr)
  ~.
\label{split}
\eea
This factorized form allows a three-step ``fast'' (in the sense of
quasi-optimal scaling) algorithm:
\ben
\item Compute all ``strengths'' $c_j$ according to the following,
  which takes $\bigO(N)$ effort:
  $$
  c_j \;:= \; e^{\frac{i\pi}{\lambda z}(x_j^2+y_j^2)}\,w_j ~, \qquad j=1,\dots,N~.
  $$
\item Evaluate
  \be
  v_k \;=\; \sum_{j=1}^N e^{\frac{-2\pi i}{\lambda z} (\xi_k x_j + \eta_k y_j)} c_j~,  \quad k=1,\dots,M
  ~,
  \label{nufft3}
  \ee
  which is precisely the task performed by the so-called 2D
  nonuniform FFT (NUFFT) of type 3 \cite{dutt,nufft3},
  a well-established algorithm with modern software
  implementations \cite{usingnfft,finufft}. This takes $\bigO(N + M + q^2\log q)$
  effort, where $q=\bigO(\fr)$ is the largest magnitude of the exponent in \eqref{nufft3}.
\item Post-multiply all outputs by their quadratic phases,
  which takes $\bigO(M)$ effort:
  $$
  \uap_k \; = \; \frac{1}{i\lambda z} e^{\frac{i\pi}{\lambda z} (\xi_k^2+\eta_k^2)} v_k ~,
  \qquad k=1,\ldots, M~.
  $$
\een

The overall cost is thus $\bigO(N + \fr^2\log \fr + M)$,
Recalling \eqref{Ngro}, this is $\bigO(\fr^2\log \fr + M)$.
The NUFFT requires a user-chosen error tolerance $\eps$ which
affects the prefactor of this run-time scaling,
but rather weakly \cite{finufft}.

\subsection{The fast algorithm for target points lying on a grid}
\label{s:t1}

Often sampling the diffracted wave on a dense regular Cartesian
grid is sufficient.
In this special case yet more speed can be gained.
Specifically, let $(\xi_k,\eta_k)$ be
the 2D grid points defined by the product of $n$-point regular 1D grids
$$
  \{-nh/2, (-n/2+1)h, \dots, -h, 0, h, \dots, (n/2-1)h\}
$$
in the $\xi$ and $\eta$ directions, $h$ being the grid spacing.
We assume $n$ is even. This grid has $M=n^2$ targets,
and (ignoring its left-most column and bottom row) is centered on the origin.
If we relabel the grid points as $(hk_1,hk_2)$
for integer indices $-n/2\le k_1,k_2 < n/2$,
and define rescaled source points
$\tilde{x}_j := (2\pi h/\lambda z) x_j$ and
$\tilde{y}_j := (2\pi h/\lambda z) y_j$,
the middle step \eqref{nufft3} can be written
  $$
  v_{k_1,k_2} \;=\; \sum_{j=1}^N e^{-i(k_1 \tilde{x}_j + k_2 \tilde{y}_j)} c_j~,  \quad
 -n/2\le k_1,k_2 < n/2~,
  $$
 which is precisely the so-called
 type 1 NUFFT \cite{dutt}
 \ch{ (also known as the ``adjoint NFFT'' \cite{usingnfft}). }
 Its nonuniform points $(\tilde{x}_j, \tilde{y}_j)$
  are only defined modulo $2\pi$, and
  may need to be ``folded'' back into a valid input domain
  such as $[-\pi,\pi)^2$.
    If the grid width is similar in size to $\Omega$, then it is easy to
    check that such a folding is only needed if $n$ is less than of order
    $\fr$, ie, the target grid under-resolves the diffracted field $u$.
    This is probably not a common use case.

    The total cost for the regular grid case is
    $\bigO(N + M \log M)$, which, recalling \eqref{Ngro} and $n=\bigO(\fr)$,
    is $\bigO(\fr^2 \log \fr)$.
    In practice, we find that for the same number $M$ of targets spanning the same domain, and the same tolerance $\eps$,
    this regular grid version is around four times faster than the
    above arbitrary target version, because the type 1 NUFFT is
    faster than the type 3 for the same space-bandwidth product \cite{finufft}.
    The generalization to rectangular target grids
    \ch{with arbitrary translations}
    is simple,
achieved through pre-phasing in step 1, and we will not present it here.

\section{Performance tests and results}
\label{s:test}

We now test the accuracy and speed of the above method
in several geometries, and compare it to two edge integral methods.
All codes are written in MATLAB R2017a, apart from FINUFFT which
is a \ch{parallel} C++ library with MEX interface.
All timings will be reported for
double-precision arithmetic on a 4-core i7-7700HQ laptop with 32GB RAM,
using 8 threads (full hyperthreading).
\ch{All codes take advantage of the multiple threads,
  either via vectorized MATLAB statements, or via OpenMP
  in FINUFFT.}
In each geometry we will first need to describe the areal quadrature
used.

\subsection{Domains defined by a simple smooth closed curve}
\label{s:kite}

We will start by setting up a simple areal quadrature
for the interior of a simple smooth closed curve.
Suppose we have good quadrature nodes
$(X_i,Y_i)\in\pO$ and weights $(W_i,V_i)$, indexed by $i=1,\dots,n$,
for {\em vector} line integrals on $\pO$,
that is, for all sufficiently smooth vector-valued functions $\ff$ on $\pO$,
\be
\int_\pO \ff(x,y) \cdot d\sss \; \approx \;
\sum_{i=1}^n \ff(X_i,Y_i) \cdot (W_i,V_i)~,
\label{vliq}
\ee
where $d\sss$ is the counter-clockwise vector line element on $\pO$.
Now fix a number of ``radial'' nodes $m$,
and let $\{\al_l\}_{l=1}^m$
be the Gauss--Legendre \cite[Ch.~19]{ATAP} nodes, and $\{\tilde{w}_l\}_{l=1}^m$
their weights, for the interval $(0,1)$.
\ch{The integral over $\Omega$ may be rewritten using a}
``dilation'' parameterization
$\iint_\Omega f \, dxdy = \int_\pO \bigl( \int_0^1 f(\al \xx) \al d\al\bigr)
\, \xx \times d\sss$,
where $\xx\in\pO$, and the 2D cross product is understood to give a scalar.
\ch{Applying the $m$-node rule on $(0,1)$ to the inner integral,
and \eqref{vliq} to the outer integral, gives the
``tensor product''}
areal quadrature for $\Omega$, with $N=nm$ nodes $(x_j,y_j)$ and weights
$w_j$, given by
\bea
(x_{l + (i-1)m},y_{l + (i-1)m}) &=& (\al_lX_i,\al_lY_i)~, \hspace{.8in} \mbox{ for }
i=1,\dots,n~, \; l=1,\dots,m ~,
\label{aqnod}
\\
w_{l + (i-1)m} &=& \al_l\tilde{w}_l(X_iV_i -Y_iW_i)
~, \qquad \mbox{ for }
i=1,\dots,n~, \; l=1,\dots,m ~.
\label{aqwei}
\eea
These nodes lie along \ch{``spokes''} connecting the origin to the
boundary nodes, as in Fig.~\ref{f:meth}(c).
If $\Omega$ is not star-shaped about the origin, then some of the
nodes lie outside $\Omega$ and some $w_j$ are negative;
however, we observe little loss of accuracy unless $\Omega$ is highly non-convex or poorly centered on the origin.

\begin{rmk}
  \ch{The above ``dilation'' method automatically creates an areal
    quadrature for $\Omega$ given only a (vector) line integral quadrature for $\pO$ and the convergence parameter $m$.
    It thus also applies to boundaries with corners or cusps, including non-symmetric starshades.
    In practice its construction time is negligible compared to that
    of the proposed NUFFT algorithm.
  }
\end{rmk}

Finally, we must build a vector line integral rule \eqref{vliq}.
The simplest case is when
$\pO$ is parameterized in a counter-clockwise sense
over $t\in[0,2\pi)$ by a smooth $2\pi$-periodic
  vector function $\xx(t) := (X(t),Y(t))$. Then
\be
\int_\pO \ff \cdot d\sss \; = \;
\int_0^{2\pi} \ff(\xx(t))) \cdot \xx'(t) dt
\; \approx \;
\sum_{i=1}^n \ff(\xx(2\pi i/n)) \cdot \frac{2\pi}{n} \xx'(2\pi i/n) ~,
\label{vliq1}
\ee
where \ch{$\xx'(t):=d\xx/dt$,
  and}
  we applied the $n$-point {\em periodic trapezoid rule} quadrature
with nodes $t = 2\pi i/n$ and equal weights $2\pi/n$.
Comparing right-hand sides of \eqref{vliq} and \eqref{vliq1},
one reads off
\be
X_i = X(2\pi i/n), \;\; Y_i = Y(2\pi i/n), \quad
W_i =  \frac{2\pi}{n} X'(2\pi i/n), \;\;
V_i =  \frac{2\pi}{n} Y'(2\pi i/n), \;\; i=1,\dots,n.
\label{ptr}
\ee

\bfi %
\ig{width=\textwidth}{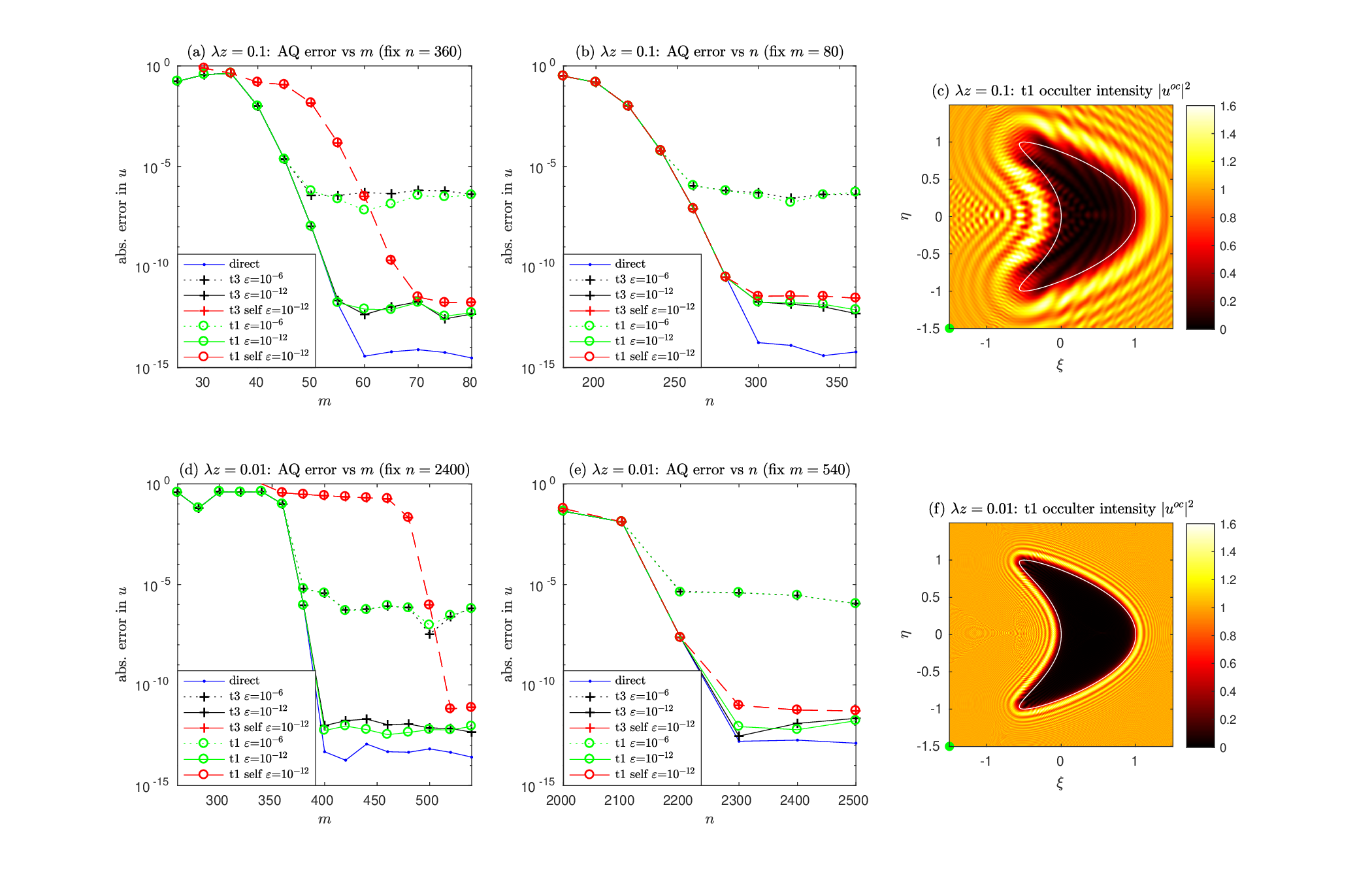}
\vspace{1ex}
\ca{
  Convergence the proposed method for the smooth kite domain
  (see Sec.~\ref{s:kite}), with respect to $m$ (``radial'' nodes),
  and $n$ (boundary nodes).
  Unless labeled ``self'', all errors are measured at the target $(-3/2,-3/2)$
  shown as a green dot in the right-most plots, relative to the
  reference line integral method (NSLI).
  Those labeled ``self'' give the maximum error over $M=10^6$ target points
  relative to their converged values.
  ``direct'' (blue) uses \eqref{faq};
  ``t3'' ($+$ signs) uses the arbitrary-target type 3 NUFFT of Sec.~\ref{s:t3}
  (random points lying in $[-3/2,3/2]^2$),
  and
  ``t1'' ($\circ$ signs) the grided target type 1 NUFFT of Sec.~\ref{s:t1}
  (grid of $n=10^3$, $nh=3$).
  Two NUFFT tolerances $\eps=10^{-6}$ and $\eps=10^{-12}$ are compared.
  The occulter Fresnel number for
  the top row (a--c) is $\fr\approx12.8$, the bottom (d--f) is $\fr\approx 128$.
}{f:kite}
\efi

\begin{table} %
{\footnotesize %
  \begin{tabular}{ll|ll|l|l|ll|l}
    $\lambda z$ & $\fr$ & $n$ (bdry) & $m$ (radial) & $M$ (targets) & method & median err & max err & CPU time \\
    \hline
    0.1 & 12.8 & 320 & --- & $10^6$, random & NSLI & --- & --- & 25.8 s  \\
    &&&&& BDWF & 2.1e-3  & 1.9e1 & 35.2 s   \\
    &&320 & 80 & &NUFFT t3 ($\eps{=}10^{-6}$) & 6.2e-8 & 1.0e-6 & 0.23 s \\
    & & & & &NUFFT t3 ($\eps{=}10^{-12}$) & 3.3e-13 & 2.8e-12 & 0.32 s \\
    & & & & $10^6$, grid & NUFFT t1 ($\eps{=}10^{-6}$) & 9.6e-9 & 8.0e-7   & 0.06 s\\
    & & & & & NUFFT t1 ($\eps{=}10^{-12}$) & 1.4e-13  & 2.6e-12 &  0.12 s\\    
    \hline
    0.01 & 128 & 2400 & --- & $10^6$, random & NSLI & --- & --- & 79 s  \\
    &&&&& BDWF & 1.1e-5 & 1.1e1 & 115 s \\
    &&2400 & 560 & &NUFFT t3 ($\eps{=}10^{-6}$) & 9.3e-8 & 4.7e-6 & 0.29 s \\
    & & & & &NUFFT t3 ($\eps{=}10^{-12}$) & 3.4e-13 & 9.5e-12 & 0.51 s \\
    & & & & $10^6$, grid & NUFFT t1 ($\eps{=}10^{-6}$) & 2.1e-8 & 4.6e-6 & 0.17 s\\
    & & & & & NUFFT t1 ($\eps{=}10^{-12}$) & 1.8e-13 & 9.6e-12 & 0.21 s\\    
    \hline
    0.001 & 1280 & 24000 & 5600 & $10^7$, random & NUFFT t3 ($\eps{=}10^{-6}$) & 5.2e-8 & 5.3e-6 & 16.2 s\\
     &&& & $10^7$, grid & NUFFT t1 ($\eps{=}10^{-6}$) & 1.9e-8 & 3.7e-6 & 10.4 s
  \end{tabular}
} %
\vspace{2ex}
\ca{\ch{Absolute} error in $\uoc$ and run times of several methods for the smooth kite occulter with converged quadrature parameters.
  The domain, upper two $\lambda z$ choices, and target region are as in Fig.~\ref{f:kite}.
  Blank entries in the table to be taken as repeated from above, and ``---''
  indicates not applicable.
  In both random and grid cases, errors are measured relative to the
  NSLI reference method (see Appendix). Timings for NSLI and BDWF
  \cite{cady12} are not listed for the grid cases since they are
  identical to the random cases.
  The last two rows are close to the largest Fresnel number $\fr$
  that the laptop
  can handle (errors were checked at only $10^4$ targets in those cases).
  See Sec.~\ref{s:kite} for other details.
}{t:kite}
\end{table}

We apply the above to build a family of areal quadratures for the kite domain
with smooth boundary
$(X(t),Y(t)) = (0.5\, \cos t + 0.5 \, \cos 2t , \, \sin t)$,
shown in Fig.~\ref{f:meth}. Its maximum radius is $R\approx 1.13$.
We then show multiple types of error convergence
for the proposed method for Fresnel diffraction in Fig.~\ref{f:kite}.
The graphs not labeled ``self'' (ie, blue, black and green) show
absolute error in $\uap$ at a single point, using
as a reference solution the (converged) NSLI edge integral method
in the Appendix.
The other ``self'' convergence (red) graphs
show the maximum error in $\uap$ over $10^6$ targets,
using the converged values themselves as a reference.
The ``direct'' (blue) simply sums \eqref{faq} without the fast algorithm.
The other graphs test two flavors of
proposed NUFFT method---arbitrary targets (t3, $+$ signs) and
gridded targets (t1, $\circ$ signs)---%
each at two different requested tolerances (6-digit and 12-digit).

In each row of panels, the first shows convergence in $m$
(``radial'' nodes), with fixed $n$ (boundary nodes), and
the second vice versa.
The right-most panels 
image, on a $M=1000^2$ point grid, the converged
intensity $|\uoc|^2$,
applying \eqref{bab} so that $\Omega$ is an occulter.
There are several observations:
\bi
\item In all cases the convergence is very sudden, as is typical with a
  spectrally-accurate quadrature rule applied to an oscillatory integrand.
\item Comparing the top to the bottom rows, where $\fr$ has increased by a factor 10,
  the converged $m$ and $n$ have each become about ten times larger.
  This matches \eqref{Ngro}.
\item The ``direct'' application of \eqref{faq} converges to 13--14 digits,
  while the NUFFT methods convergence bottoms out at within one digit of the
  requested $\eps$, as expected.
\item 
  The ``self'' convergence (red graphs) in $n$ (panels b, e) occurs
  in tandem with the independent error at the single test point $(-3/2,-3/2)$.
  However, for $m$ (panels a, d) this is not quite true, even though the
  test point is the point in the target domain $[-3/2,3/2]^2$ maximizing
  the maximum source-target separation $r$. This is due to
  particulars of the angular variation in node density, and reminds one
  that convergence must be tested at {\em all} target points.
\ei

Table~\ref{t:kite} reports CPU timings and errors for these same tasks
at converged $n$ and $m$ quadrature parameters.
(The areal quadrature generation is not included, but was found to be
negligible.)
NSLI (known to achieve 13--14 digits with these parameters)
is used as the reference for all errors.
A state of the art edge integral code, BDWF \cite{cady12}
(as available in SISTER\cite{SISTER}, and documented in
\cite{fresnaqgit}) is also
tested with the same $n$: while its median errors are as expected
from its use of a 2$^\tbox{nd}$-order accurate midpoint
rule, its maximum errors are larger than 1. These
huge errors appear to be confined
to a few target points very near $\pO$, the geometric shadow edge.
The main conclusions from Table~\ref{t:kite} are then:
\bi
\item In this setting, the proposed NUFFT based methods
  are $100$ to $300$ times faster than the edge integral methods (for arbitrary
  targets), or $400$ to $600$ times faster (for gridded targets).
\item The proposed methods {\em robustly} (uniformly at all targets)
  achieve close to the requested error.
\item At the largest $\lambda z$ ($\fr \approx 12.8$), the t1
  achieves $\sim 10^7$ targets/second, and the t3 achieves $\sim 4\times 10^6$ targets/second, with only a weak dependence on tolerance.
  These are only slightly slower for $\fr$ ten times larger.
\item At the smallest $\lambda z=10^{-3}$ ($\fr \approx 1280$)
  the asymptotic $\bigO(\fr^2 \log \fr)$ cost has started to dominate.
  Note that 134 million nodes are being mapped to 10 million targets in
  around 10 seconds.
\ei

Since under the hood the NUFFT uses FFTs, the reader might worry
about their RAM usage.
It is very mild: for t1 (gridded) cases the FFT size is $5/4$ times
(for $\eps\ge 10^{-9}$, or twice otherwise)
the requested grid size, in each dimension.
For t3, FFT dimensions scale like $\fr$:
for the smallest $\fr\approx 12.8$, the FFT is a tiny $216\times 288$.
This is to be compared with
the $32768\times32768$ FFT needed for a sub-pixel sampling
method to reach around 6-digit accuracy in $u$
in various tests \cite{harness18} at similar $\fr$.
In the penultimate row of the table $\fr$ is $10^2$ times larger,
yet the FFT is only $7500\times 9600$ (similar to the maximum $r^2/\lambda z \approx 8900$ zones),
and total RAM usage is 17 GB,
about the largest the laptop can handle.

\ch{
  Another natural question is: does the method suffer at smaller $\fr$
  than tested above? It does not: node numbers and CPU times only get smaller.
  In fact, by expanding the target grid in proportion to
  $z \to \infty$, the
  Fraunhofer limit is reached in a stable fashion
  (here the first factor in \eqref{split} should be discarded, while the third
  factor tends to unity).
}

\bfi  %
\centering \ig{width=.7\textwidth}{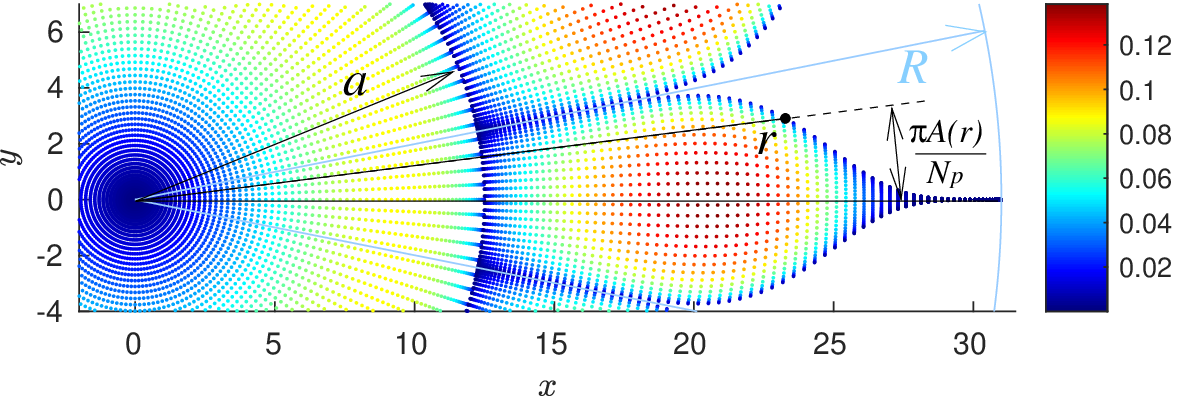}
\ca{Idealized starshade geometry (zoom for clarity).
  $A(r)$ is the apodization profile controlling petal angular width,
  here illustrated with the HG (offset hyper-Gaussian \cite{cash11})
  with $N_p=16$ petals.
  The colored dots show areal quadrature nodes $(x_j,y_j)$
  with weights $w_j$ indicated by color as in the colorbar.
  $m=80$ Gauss radii cover the petal length, with
  $n_p=30$ Gauss nodes covering the petal angular width at each radius.
  }{f:starnodes}
\efi

\subsection{Application to starshade modeling}
\label{s:star}

Idealized starshades are described \cite{vanderbei07,cash11,cady12}
by a radial apodization
function $A(r)$, where (in this section only) we use $(r,\theta)$ as
occulter-plane polar coordinates about the origin.
$A(r)$ is $1$ (indicating a fully blocking disc) for $r<a$,
and drops in a carefully optimized
fashion in $r\in[a,R]$ to close to zero at $R$, the maximum occulter radius,
and identically $0$ beyond this.
Apodization over $[a,R]$ is realized via $N_p$ identical binary
petals, each of whose angular width at radius $r$ is $2\pi A(r)/N_p$.
Let the function $P(\alpha)$ denote
$\alpha + 2\pi n$
  for the unique $n\in\ZZ$ such that $\alpha + 2\pi n \in [-\pi,\pi)$,
    a common definition of the principal value of an angle.
Then the occulter is the ``flower'' shape,
\be
\Omega \;=\;
\bigl\{(r,\theta) : \;\; 0 \le r \le R, \;\; P(N_p\theta) \in (-\pi A(r), \pi A(r)) \; \bigr \}~.
\label{Omega}
\ee
See Fig.~\ref{f:starnodes}.
Note that published $A(r)$ designs are
{\em discontinuous} at $a$ (indicating a gap between
petals), and at $R$ (petal tips have finite width).
In early ``analytic'' designs these discontinuities were
required to be no larger than about $10^{-5}$ in size
to minimize Arago-spot-style
diffraction into the deep shadow \cite[\S4.3]{cash11},
but designs generated by optimization
over a $\lambda z$ band \cite{vanderbei07,SISTER}
have much larger gap and tip discontinuities, of order $10^{-2}$ to $10^{-3}$,
whose Arago effect is apparently cancelled out \ch{over the band} by distributed
``ripples'' \cite[\S5]{cash11} in $A(r)$.

Recall that the task is simply to evaluate \eqref{fres} and \eqref{bab} with
errors in $\uoc$ no worse than $10^{-6}$.
To apply the proposed method, we build a high-order areal quadrature as follows.
Since $A(r)$ is discontinuous at $r=a$,
we split $\Omega$ into the disc of radius $a$ plus each of $N_p$ petals.
Our disc quadrature simply applies
\eqref{aqnod}--\eqref{aqwei} to the uniform $n_d$-node line integral on its
boundary, that is, \eqref{ptr} applied to the parameterization
$(X(t),Y(t)) = (a\cos t, a \sin t)$. We use $m_d$ radial nodes.
We then cover each petal by $m$
nodes \ch{to handle the (outer) radial integral, then handle the
(inner) arc integral at each of their node radii by $n_p$
angular nodes; see} Fig.~\ref{f:starnodes}.
Specifically, let $\{r_l, \hat{w}_l\}_{l=1}^m$
be 1D Gauss--Legendre nodes and weights for the (outer) radial integral
over $(a,R)$. Similarly, let
$\{t_i, \hat{\omega}_i\}_{i=1}^{n_p}$ be nodes and weights for
the fixed interval $[-\pi/N_p,\pi/N_p]$.
Then, \ch{recalling the area element $r \,dr\, d\theta$,}
the resulting areal nodes (in Cartesians) and weights \ch{covering}
one petal are
\bea
(x_{l+(i-1)m},y_{l+(i-1)m}) &=& \bigl(
  r_l \cos(A(r_l)t_i),r_l \sin(A(r_l)t_i)
  \bigr)
  ~, \quad \mbox{ for }  i=1,\dots,n_p~, \; l=1,\dots,m ~,
\nonumber
\\
w_{l+(i-1)m} &=& r_l \hat{w}_l  A(r_i) \hat{\omega}_i
~, \hspace{1.48in} \mbox{ for }  i=1,\dots,n_p~, \; l=1,\dots,m ~.
\nonumber
\eea
Other petals are obtained by rotation by multiples of $2\pi/N_p$.
The total number of nodes is then $N = n_d m_d + N_p n_pm$.
We will fix $m_d = m$ and $n_d \approx 0.3 N_p n_p$,
leaving two (petal) convergence parameters $m$ and $n_p$.
\ch{ In Sec.~\ref{s:conc} we discuss applying this to perturbed
  (non-ideal) starshades.}

\bfi  %
(a)\raisebox{-2.5in}{\ig{width=.47\textwidth}{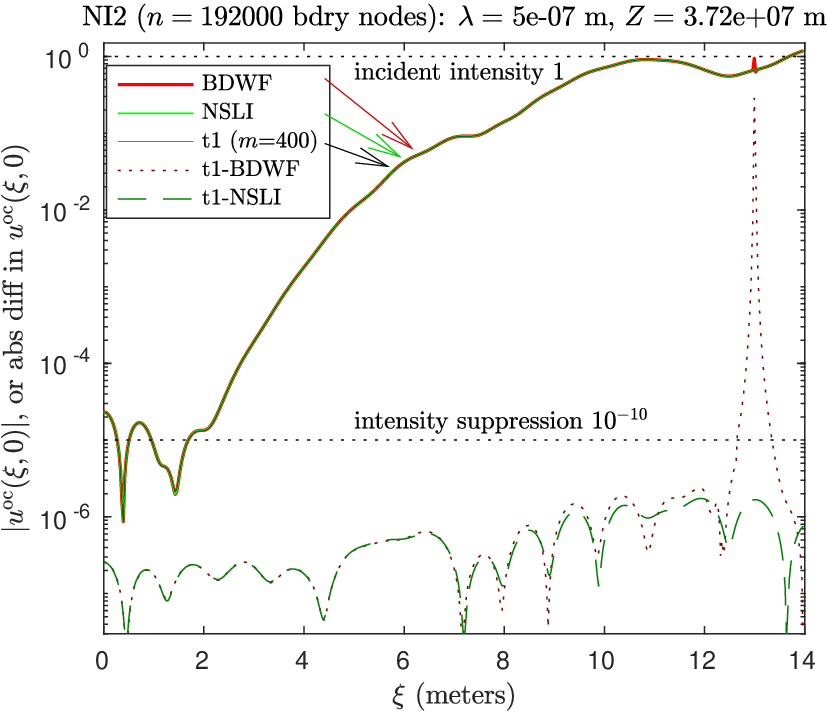}}
\hfill
(b)\raisebox{-2.5in}{\ig{width=.46\textwidth}{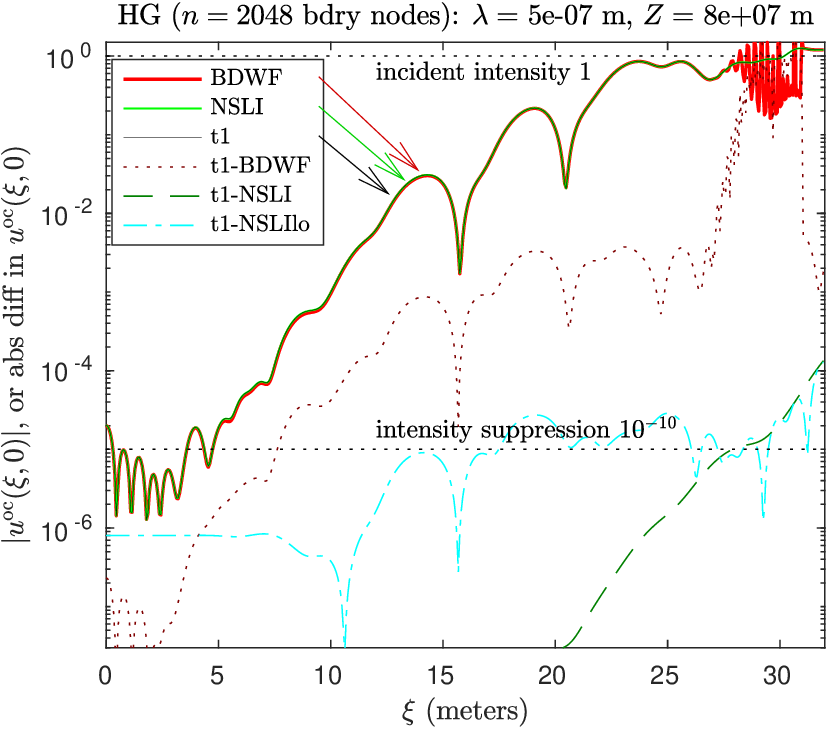}}
\vspace{1ex}
\ca{Validation of diffracted $\uoc$ along a radial slice for two starshade designs:
  NI2 (optimized function with ripples \cite{shaklan17}) and HG
  (offset hyper-Gaussian analytic function \cite{cash11}).
  Both have $N_p=16$ petals.
  The proposed NUFFT t1 method (black line) is compared against
  edge-integral methods BDWF (red) and NSLI (green).
  Labels such as ``t1-BDWF''
  indicate the absolute difference in $\uoc$ between two methods.
  BDWF and NSLI use the same boundary nodes, except for ``NSLIlo''
  which uses $10\times$ the boundary nodes and 2$^\tbox{nd}$-order weights.
  For details see Sec.~\ref{s:star}.
  }{f:valid}
\efi

\subsubsection{Accuracy validation}
\label{s:valid}

We compare in Fig.~\ref{f:valid}
the wave amplitudes along a radial slice
computed by three methods,
for two designs of starshade:
``NI2'' (a small occulter with rippled profile optimized for a blue-green $\lambda$ range \cite{shaklan17}), and ``HG''
(a large occulter with analytic ``offset hyper-Gaussian'' profile
\cite{cash11}).
We choose $\lambda$ within their designed wavelength windows.
The parameters used, and some CPU timings, are listed in Table~\ref{t:star}.
Since they have different geometry descriptions,
we treat the two designs in turn.

{\bf NI2.}
The optimized profile $A(r)$ is available in the SISTER package \cite{SISTER}
in the form of $2462$
equispaced samples covering the petal radius range $[a,R] = [5,13]$ m.
From these, we use piecewise cubic splines to interpolate $A$ at $m$
radial Gauss nodes in $[a,R]$.
Since $A''(r)$ appears to have at least 13 ``bang-bang'' type discontinuities
(the discrete 2$^\tbox{nd}$ derivative mostly takes values
$\pm\sigma$, for some constant $\sigma$, or $0$), this {\em necessarily} limits accuracy
to around 6-7 digits.
By a convergence study we
found that $m=400$, and $n_p=40$ nodes across each petal, were sufficient
for areal quadrature to match this accuracy.
For BDWF we used the $n=192000$ boundary nodes as given and used in SISTER.
These have $6000$ nodes per petal edge, but no nodes covering the
inter-petal gaps, or tips (each of which is $0.03$ m wide).
For NSLI we used an $n$-node vector line integral quadrature matching the
2$^\tbox{nd}$-order accurate midpoint rule in BDWF (this match was needed to
accurately handle the wide tips with a single segment).
Fig.~\ref{f:valid}(a) shows that the proposed NUFFT t1 method matches both
of these edge integral methods to
around $3\times 10^{-7}$ in $\uoc$ in the shadow region where
$|\uoc| \le 2\times 10^{-5}$.
NSLI agrees with t1 to around 6-digits everywhere.
However, at $\xi\approx 13$ m, the error of BDWF spikes to $\bigO(1)$ as
$(\xi,0)$ approaches $\pO$.

\begin{rmk}
Since $z/\lambda \sim 10^{14}$, overall phase is meaningless, thus
we fit the phase of BDWF to the other two methods at a single target.
We then quote absolute differences in complex $\uoc$.
This is a more predictable metric than the
error in intensity $|\uoc|^2$, which is affected by local intensity.
\end{rmk}

{\bf HG.}
Here the profile is analytically known
\cite{cash11}
: $A(r) = e^{-[(r-a)/b]^p}$ in $[a,R]$,
where $a=b=12.5$ m, the maximum radius is $R=31$ m, and $p=6$.
A convergence study shows that only $m=60$ and $n_p=30$
are needed, giving an areal quadrature of $N=37440$ nodes.
For NSLI we used $A(r)$ and $A'(r)$ to generate a high-order line integral
quadrature
using the same $m=60$ radii per petal, plus four Gauss nodes across each
gap and tip, giving $n=2048$ in total
(see {\tt starshadeliquad} in our repository\cite{fresnaqgit}).
We also fed these boundary nodes (but of course not their high-order weights)
to BDWF.
Fig.~\ref{f:valid}(b)
shows that the three methods again agree to the desired accuracy
almost everywhere,
apart from BDWF near $\pO$ where again its errors hit $\bigO(1)$.

\begin{rmk}
  For HG with $m=60$ the errors of BDWF
  are summarized by 2-3 digits of {\em relative} accuracy overall,
  giving 6-7 digits of absolute $\uoc$ accuracy in the deep shadow.
Yet NSLI, if fed the low-order midpoint rule used inside BDWF, gives only
4-digit {\em absolute} accuracy, thus is useless in deep shadow.
Fig.~\ref{f:valid}(b) thus also explores (dash-dot line)
the errors of NSLI with this low-order rule
and the larger $m=600$: the absolute error now
bottoms out at a useful $10^{-6}$.
This highlights an advantage of BDWF over plain NSLI when {\em deep shadows
are modeled with poor quadrature}, a subtle point explained in
Remark~\ref{r:shadfix}.
\label{r:shadacc}
\end{rmk}

\begin{table} %
{\footnotesize %
\quad
  \begin{tabular}{llll|ll|l|l|l}
    design & $\lambda$ (m) & $z$ (m) & $\fr$ & $m$ (petal) & total nodes & $M$ (targets) &method & CPU time \\
    \hline
    NI2 & 5e-7 & 3.72e7 & 9.1 &
    6000 & $n{=}192000$ & $10^6$, grid & BDWF & 5361 s \\
    &&&&
    400 & $N{=}499200$ && NUFFT t1 ($\eps{=}10^{-8}$) & 0.076 s \\
    \hline
    HG & 5e-7 & 8e7 & 24 &
    60 & $n{=}2048$ & $10^6$, grid & BDWF & 80.5 s \\
    &&&&
    60 & $N{=}37440$ && NUFFT t1 ($\eps{=}10^{-8}$) & 0.042 s \\
    \hline
  \end{tabular}
} %
\vspace{2ex}
\ca{Parameters and CPU times for the proposed NUFFT t1 and the BDWF edge-integral
  to complete the same diffraction
  tasks, for two starshades.
See Fig.~\ref{f:valid} for
  comparisons of their answers.
  The Fresnel number $\fr$ uses the maximum radius $R$ in \eqref{fn}.
  $N$ is the number of areal quadrature nodes, while $n$ the number of
  boundary nodes.
}{t:star}
\end{table}

\bfi  %
\ig{width=.47\textwidth}{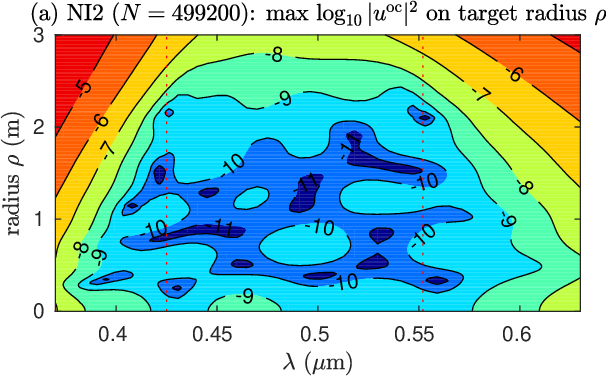}
\hfill
\ig{width=.51\textwidth}{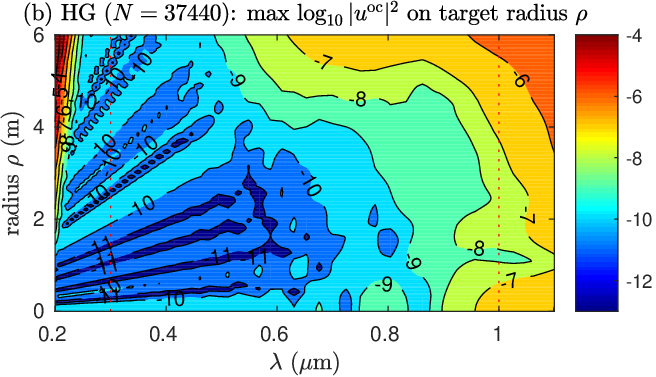}
\ca{Intensity (on $\log_{10}$ scale indicated on the right)
  as a function of wavelength and target radius $\rho$ from the center,
  for the two starshade designs (NI2 and HG) of Fig.~\ref{f:valid}.
  At each of 200 $\rho$ values, the maximum over $300$ angles is taken.
  The indicent intensity is 1. The NUFFT t3 method is used.
  Vertical dotted lines show the designed $\lambda$ range.
}{f:lambdas}
\efi

\subsubsection{Solution speed}
\label{s:speed}

Table~\ref{t:star} presents CPU times for the above
converged experiments on gridded targets (we omit NSLI times since they are similar to BDWF).
\ch{The time to construct the areal quadrature is not included, for two
  reasons:
  i) it is never more than half the NUFFT method run-time, and ii)
  it would be amortized away over runs at multiple wavelengths.}
We make the following observations:
\bi
\item For NI2 our proposal is around $70000\times$ faster, and for
  HG around $2000\times$ faster, than a state-of-the-art edge integral method.
\item
  By reinterpolation of the NI2 profile,
  and code changes to use a high-order quadrature, BDWF
  could probably be sped up by a factor of 15.
  BDWF is found also to gain a factor of about two when multiple
  $\lambda$ are needed. Neither factor impacts the conclusions much.
\item Since the number $N$ of areal nodes is smaller than
  the large number $M$ of targets, the cost of the NUFFT method
  is almost independent of $N$, hence of the starshade complexity, or $\fr$.
\ei

We now turn to arbitrary targets.
In Fig.~\ref{f:lambdas}
the intensity suppression of the two starshade designs are
studied, sweeping 50 wavelengths, and taking the maximum intensity
$|\uoc|^2$ over circles of varying radius $\rho$ ($M=60000$ targets
at each $\lambda$), using the proposed NUFFT t3 method.
The narrow-band nature of NI2, and deterioration of HG above
$0.8\,\mu$m, are apparent.
The entire calculation for both starshades, \ch{including
quadrature generation}, totals 6 seconds on the laptop.

Finally, we report initial timing results upon having
(rather crudely) inserted the NUFFT t3 method in place of BDWF
within the SISTER code base \cite{SISTER}.
(See {\tt sister\_mods} in our repository \cite{fresnaqgit}.)
Running a standard SISTER ``PSF basis'' generation task
for the non-spinning NI2 starshade,
14 wavelengths are needed covering $[0.425, 0.555]\, \mu$m,
at each of which
$M=806144$ targets are needed.
(Targets are organized into $16\times 16$ telescope pupil grids,
translated to $3149$ different $(\xi,\eta)$ centers covering
a sector with angle $2\pi/N_p$.)
We had to reorganize the loop ordering, since BDWF was called separately
for each pupil while grouping wavelengths together for speed,
whereas the NUFFT t3 is most efficient with a {\em single call to all targets}
at each wavelength.

The original SISTER run time was an estimated 6.5 hours
(since BDWF gets about $6\times 10^7$
node-target pairs per second).
The NUFFT t3 method, using parameters as in the previous section,
and \ch{including quadrature generation},
took 2.6 seconds. The acceleration is thus about $10000\times$.

\begin{rmk}
  \ch{The above shows that the proposed algorithm excels in efficiency when the number of
  desired targets $M$ is large, resolving a region similar in size to the occulter. Since its cost is close to $\bigO(N+M)$,
  dropping $M$ does not reduce run-time much: $N$ then dominates,
  and the {\em relative} speed over edge integral methods drops in
  proportion.
  The natural question is: what is the crossover $M$ such that there is
  no advantage? For the NI2 starshade, since BDWF gives about
  $250$ targets/s, the answer is as small as $M\approx 20$ for t1 and $50$ for t3.
  Thus whenever the user needs more targets than this, the NUFFT wins.}
  \end{rmk}

\bfi  %
\ig{width=.52\textwidth}{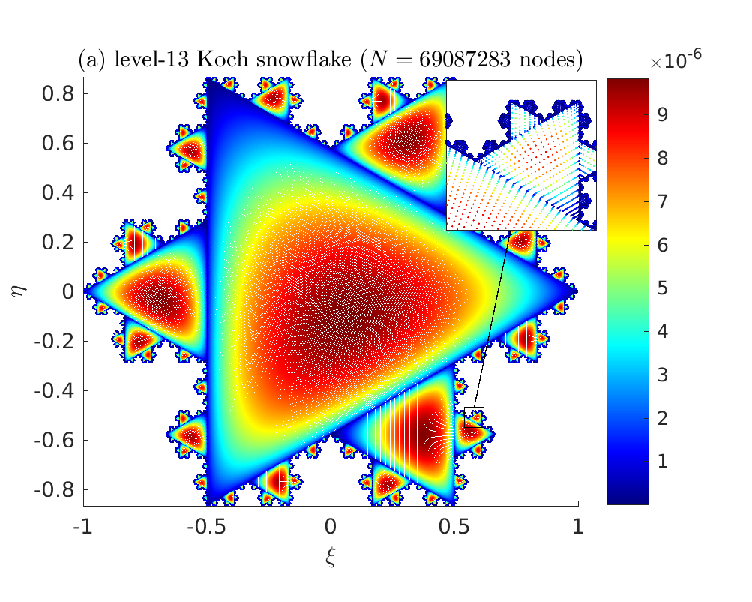}
\ig{width=.47\textwidth}{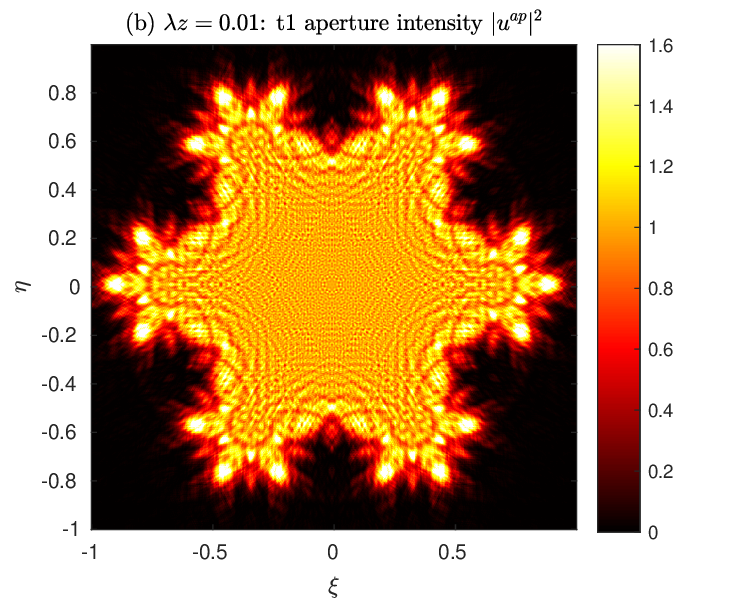}
\ca{Koch fractal aperture diffraction example from Sec.~\ref{s:koch}.
  (a) shows the areal quadrature constructed by a union of about 67 million
  triangles.
  The color of each node $(x_j,y_j)$ indicates its weight $w_j$ using the
  scale on the right.
  The inset shows a zoom into the region shown, resolving individual nodes.
  (b) shows intensity (on $\log_{10}$ scale indicated on the right)
  computed on a million-point grid by the NUFFT t1 method in under 5 seconds.
}{f:koch}
\efi

\subsection{A complicated domain}
\label{s:koch}

As a final example we compute Fresnel diffraction for an
aperture with fractal boundary, specifically the standard Koch snowflake with
maximum radius $R=1$.
Let $\Omega_0$ denote the equilateral triangle with side length $\sqrt{3}$,
$\Omega_1$ its union with
the three triangles of side $\sqrt{3}/3$,
$\Omega_2$ the union of $\Omega_1$ with the 12 triangles of side $\sqrt{3}/9$,
etc, so that $\Omega_L$ is the level-$L$ construction; see Fig.~\ref{f:koch}(a).
To reach level $L=13$, $n_\tbox{tri} = 1+3\sum_{k=0}^{L-1}4^k = 67108864$
triangles are needed.
To build an areal quadrature, the integral over each triangle is approximated
by a simple $p\times p$ node product Gauss--Legendre quadrature,
by translating one vertex to the origin then
applying \eqref{aqnod}--\eqref{aqwei} to the discretized line
integral connecting the other two vertices (see inset of figure).
To accurately handle the oscillatory integrand as in \eqref{Ngro},
the node spacing should not exceed $\bigO(1/\fr)$;
thus we designed a heuristic choice of $p$ that varied from $217$ for the
largest triangle to $p=1$ at levels $L\ge10$, and
checked $p$-convergence for Fresnel integrals for $\lambda z \ge 0.01$.
For $\Omega_{13}$, the resulting total node number $N$ is about 69 million,
requiring about 4 minutes to build in our simple implementation.

The NUFFT t1 method with $\eps=10^{-6}$
is then applied to this areal quadrature, to
resolve the diffracted field for $\lambda z= 0.01$
on a grid of $10^6$ target points, giving Fig.~\ref{f:koch}(b).
For each new $\lambda z \ge 0.01$ this takes 4.6 seconds.
Since $\Omega_{13}$ has $3\cdot 4^{13}\approx 2\times 10^8$ edges,
an edge integral method of similar accuracy is estimated to be around
$10^5$ to $10^6\times$ slower.
We have checked (via $p$-convergence at each level) that \ch{this}
$u$ computed for $\Omega_{13}$ has at least 6-digit accuracy.

However, we may also interpret the calculation as
an approximation to one
for the {\em limit domain} $\Omega_\infty$ with
true fractal boundary.
Since the smallest triangles in $\Omega_{13}$ have side
$1.1\times 10^{-6} \ll \lambda z$, ie, much smaller than any Fresnel zone,
we are well into the regime of Richardson extrapolation in $L$,
with differences from the limit scaling like $(4/9)^L$.
The largest absolute change in $u$ on the grid in going from $\Omega_{12}$ to
$\Omega_{13}$ was $1.6\times 10^{-4}$,
thus, by extrapolation, the largest change in $u$ between
$\Omega_{13}$ and the limiting domain $\Omega_\infty$
is around $4/5$ of this.
Thus we may quote uniformly
about 4-digit accuracy for diffraction from the limit fractal domain.
(By applying Richardson to a sequence, one could in fact \ch{recover}
many more digits
without much extra effort.)

\section{Conclusion and discussion}
\label{s:conc}

We have explained a fast algorithm for Fresnel diffraction from binary aperture
or occulter shapes,
which achieves high accuracy via flexible areal
quadrature schemes over the planar domain,
yet speeds close to
regular-grid FFT propagation methods, via the nonuniform FFT.
Extensive tests of error convergence and CPU timings
show between 2 and 5 orders of magnitude acceleration over
edge integral methods at comparable accuracies.
Thus, at moderate Fresnel numbers $\fr \le 10^2$
one can evaluate accurate diffraction fields from complicated shapes
such as starshades
almost instantaneously (0.1 seconds for a grid of $10^6$ targets).
For higher $\fr$, the $\bigO(\fr^2 \log \fr)$ cost starts to dominate.
RAM usage starts to become a limitation only at $\fr\sim 10^3$.
Along the way, we have reformulated edge integrals as a non-singular
line integral that is numerically robust (Appendix~A).

Although we did not exploit it,
the proposed NUFFT method can trivially include \ch{smooth} source-plane
phase/amplitude variations
while retaining accurate edge diffraction,
\ch{simply by multiplying $w_j$ by the source term at each areal node,
  in step 1 of Sec.~\ref{s:t3}.}
This is impossible for traditional edge integral methods,
but we note exciting recent progress on including low-frequency
phase variations with edge integrals \cite[Sec.~6]{harness18}.

Our findings highlight the importance of high-order accurate quadratures,
both for edge integrals and, crucially, planar integrals.
We have shown how the latter may be constructed
for three classes of domain $\Omega$
(\ch{those with an existing boundary quadrature}, ideal starshades, and unions of triangles).
Such construction on a case-by-case basis is possible,
at least up to the tolerance with which a description of $\pO$ is available.
Yet to automate this procedure for a requested accuracy and $\fr$, given ``any''
$\Omega$, raises 2D geometry representation and meshing issues common
throughout engineering and scientific computing.
This is a huge topic with many available tools.
Given the benefits that we show in optical \ch{simulation}, automated high-order
areal quadratures from CAD formats should be a useful future project.

\subsection{\ch{Application to non-ideal (perturbed) starshades}}

\ch{%
Since the proposed method does not exploit symmetry, it could
vastly accelerate Monte Carlo studies of optical stability
under the various types of realistic shape perturbations.
The latter include manufacturing tolerances, in-flight misalignment, thermal distortions, and damage over time.
The topic is complicated by %
geometry descriptions, statistical correlations, and averaging
due to spinning \cite{shaklan17}.
Numerical study of this is beyond the scope of this initial work.
Yet, recall that once an {\em areal quadrature} exists for the desired shape $\Omega$, the proposed method is very simple.
We explain three ways that such a quadrature can be generated:
\ben
\item
  Sec.~\ref{s:kite} showed how to automatically build an areal quadrature
  from an existing boundary quadrature for $\pO$, as in Fig.~\ref{f:meth}(c).
  In this way the method may immediately be applied to
  a set of boundary nodes describing a perturbed starshade.
  This will not be as efficient as the quadratures of Sec.~\ref{s:star}:
  e.g.\ for NI2, using the rather large supplied $n=192000$ would give
  $N \sim 10^8$, giving CPU times
  around $10$ s. Yet this is still 2--3 orders of magnitude faster than edge integrals in the case of $M=10^6$ targets.
\item
  For ideal rigid petals that are misaligned, a deformed areal quadrature may be built
  as follows: apply rigid motion to the nodes for each petal, without changing their weights, then add new quadratures covering the
  new (signed) areas which connect the base arc of each petal with its ``root'' arc on the ideal disc. This would only add a few nodes to $N$, hence preserve the $0.1$ s
  CPU times quoted.
\item
  A general non-ideal shape may be written as the ideal shape plus a narrow (signed)
  ``ribbon'' domain in the neighborhood of the boundary $\pO$. If its width is everywhere smaller than a Fresnel zone, as expected for realistic perturbations, the ribbon contribution is well approximated by a line integral on $\pO$. The resulting quadrature is that of Sec.~\ref{s:star} plus a smaller number of boundary nodes, preserving $0.1$ s CPU times.
  \een
  Which of the above methods, or whether another method,
  proves best in practice will depend on
  the number of runs required and spatial details of the perturbation.
}%

\subsection{Other extensions}

\ch{ Beyond the geometry-handling extensions discussed above,}
fruitful future directions include:
\bi
\item Further acceleration is probable by drop-in
  replacement of FINUFFT by a GPU library. %
\item \ch{In this initial work, Gauss--Legendre rules were used for
  all 1D quadratures. However, since the Fresnel
  integrands are oscillatory
  and band-limited, it is probable that substituting high-order 1D rules with quasi-uniform nodes \cite{alpert,halequad} will allow $N$ to be reduced for
  the same accuracy.
  Asymptotically at high $\fr$, one would expect a reduction by up to a factor $(\pi/2)^2\approx 2.5$.}
  
\item It seems that a NUFFT replacing the first FFT in method (a)(i)
  from the Introduction could enable
  efficient new ``angular spectrum of plane waves''
  \cite[\S3.10]{goodman96} (one-way Helmholtz) methods for
  binary aperture diffraction beyond the Fresnel regime.
\ei

\section*{Acknowledgments}

This work benefited from the input of the anonymous reviewers, and
discussions with
David Spergel, Robert Vanderbei, Stuart Shaklan, and Eric Cady.
The Flatiron Institute is a division of the Simons Foundation.

\appendix
\section{Fresnel edge integral methods---equivalence and desingularization}

\subsection{Equivalence of edge integral methods for planar waves and apertures}

Dauger \cite{dauger96}
noticed that, by
using polar coordinates about the target $(\xi,\eta)$,
ie, $x = \xi + r\cos\theta$ and $y = \eta + r\sin\theta$,
the Fresnel aperture integral \eqref{fres}
may be analytically integrated in $r$, for each $\theta$, as follows.
Assuming that the target is in $\Omega$ (geometric shadow),
furthermore that $\Omega$ is strictly
star-shaped about this target, \eqref{fres} becomes
\be
\uap(\xi,\eta)\; = \;
\frac{1}{i\lambda z} 
\int_0^{2\pi} \int_0^{R(\theta)} e^{\frac{i\pi}{\lambda z}r^2} r dr \, d\theta
\; = \;
\frac{-1}{2\pi} \int_0^{2\pi} [e^{\frac{i\pi}{\lambda z}R(\theta)^2} - 1] \,d\theta
\label{dauger}
\ee
where $R(\theta)$ is the distance $r$ where the in-plane ray
launched at angle $\theta$ from the target exits $\Omega$.
If these conditions are broken, $R(\theta)$ becomes multi-valued.
For targets outside $\Omega$, the second term in square brackets
must be replaced by one similar to the first but
involving the $r$ value where the ray first enters $\Omega$.
The resulting numerical method is cumbersome for a general $\Omega$
because much effort is spent finding, and tracking as a function of $\theta$,
these multiple ray intersection points \cite{dauger96}.

It is simpler to reformulate \eqref{fres} in terms of a line
integral over $\pO$. Cash \cite[(46)]{cash11} provides such a formula but
no rigorous derivation.
To remedy this, fixing the target, we write $\rr := (x-\xi,y-\eta)$,
hence $r^2 = \|\rr\|^2$, and define the 2D vector field
\be
\FF(x,y) %
\;:=\;
\frac{-1}{2\pi}\frac{\rr}{r^2} e^{\frac{i\pi}{\lambda z}r^2} ~, \qquad (x,y) \neq (\xi,\eta)~.
\label{F}
\ee
After cancelling terms, its divergence is found to be simply
the Fresnel integrand from \eqref{fres}
minus the unit 2D delta distribution at the target
(the latter can be proven by excluding a small disk of radius $r\to0$ about
the target). That is,
\be
\nabla \cdot \FF(x,y) \;=\; \frac{1}{i\lambda z} e^{\frac{i\pi}{\lambda z}r^2}
 - \delta(\rr)~.
\label{divF}
\ee
Applying the divergence theorem in $\Omega$, with $\mbf{n}$ the unit outward normal, and, as before, the 2D cross product taken to be a scalar,
$$
\iint_\Omega \nabla \cdot \FF \, dxdy
=
\int_{\pO} \FF \cdot \mbf{n}\, ds =
\int_{\pO} \FF \times d\sss =
\frac{-1}{2\pi}\int_{\pO} e^{\frac{i\pi}{\lambda z}r^2} \frac{\rr \times d\sss}{r^2}
~.
$$
Substituting \eqref{divF} and recalling \eqref{fres} gives
the line integral formula
\be
\uap(\xi,\eta) =
\frac{-1}{2\pi}\int_{\pO} \! e^{\frac{i\pi}{\lambda z}r^2} \frac{\rr \times d\sss}{r^2} +
\uapg(\xi,\eta),
\quad \mbox{ where }\;\;
\uapg(\xi,\eta) := \left\{\begin{array}{ll}
1, & (\xi,\eta)\in\Omega
\\
0, & \mbox{otherwise}
\end{array}\right.
\label{li}
\ee
This is easily seen to be equivalent to Dauger's formulae using the facts:
i) $d\theta = (\rr \times d\sss) / r^2$,
ii) $\uapg(\xi,\eta) = \int_0^{2\pi} d\theta / 2\pi$, and
iii) the multiple values of $R(\theta)$ correspond to $\theta$ folding
back as $\pO$ is traversed.

We now show that, within the Fresnel approximation,
the Miyamoto--Wolf \cite[Eqs.~(5.1), (5.5)]{miyamoto62}
boundary diffraction wave (BDW) formulation used
by Cady \cite{cady12} is also equivalent to the above.
Given $u(x,y)$, the plane incident wave at the aperture with unit
direction vector $\mbf{p}$, this states (noting that our $\uap$ definition
excludes the phase of plane $z$-propagation),
\be
\uap(\xi,\eta) \;=\; \frac{1}{4\pi}
e^{-2\pi i z/\lambda}
\int_{\pO} u(x,y) \frac{e^{2\pi i \rho/\lambda}}{\rho}
\frac{\hat{\bm\rho}\times \mbf{p} \cdot d\sss}{1 + \hat{\bm\rho}\cdot\mbf{p}}
\;+\; \uapg(\xi,\eta)~.
\label{bdw}
\ee
Here we
recall that $\rho = \sqrt{r^2 + z^2}$ is the target-source 3D distance,
and define $\hat{\bm\rho}$ to be the unit vector pointing from target to source
($\rho$ is notated as $s$ in standard references, but we reserve the latter for
arclength).
Since we are concerned with planar incidence, $u(x,y)\equiv1$
and $\mbf{p}=(0,0,1)$.
Since $r\ll z$ is implicit in \eqref{fres},
we insert the leading-order small-angle approximations
$\rho\approx z$, \;
$\hat{\bm\rho}\times \mbf{p} \cdot d\sss \approx (\rr \times d\sss)/z$, \;
$1 + \hat{\bm\rho}\cdot\mbf{p} \approx r^2/(2z^2)$,
and the usual Fresnel approximation
$e^{2\pi i \rho/\lambda} \approx e^{2\pi i z/\lambda} e^{\frac{i\pi}{\lambda z}r^2}$.
The result is precisely \eqref{li}.
Thus all three edge formulations are equivalent.

However, it is worth noting that BDW \eqref{bdw}
and some formulae in Dubra--Ferrari \cite{dubra99} have a wider range of
validity than \eqref{fres}, Dauger's formulae, or \eqref{li},
since they allow out-of-plane apertures and more general incident waves.

\subsection{A robust non-singular line integral (NSLI) formulation}

To our knowledge all edge integral numerical codes
use formulae shown
in the previous section
to be equivalent to \eqref{li},
and are thus well known to be plagued by two serious problems:
\cite{dauger96,dubra99,cash11,cady12,harness18}
\ben
\item targets must be labeled as being inside or outside of $\Omega$ in a
  robust fashion, no matter how close they are to $\pO$,
  otherwise $\bigO(1)$ errors result,
and
\item when the target approaches $\pO$, the integrand
on $\pO$ becomes nearly singular, requiring increasingly refined
quadrature near the target to retain accuracy.
(Dauger's $\theta$-parameterization
conceals this, but does not remove the difficulty, since
$R(\theta)$ changes arbitrarily rapidly.)
\een
For instance, in Sec.~\ref{s:kite} and \ref{s:valid} we saw that the BDWF code
loses all accuracy near to $\pO$. %
However, once it is realized that the two problems are in fact
facets of the same phenomenon, they can be made to ``cancel out''.

This works as follows. It is well known \cite[(6.23)]{LIE}
(or combining facts i) and ii) above), that
\be
\uapg(\xi,\eta) \;=\; \frac{1}{2\pi}\int_\pO \frac{\rr \times d\sss}{r^2}~.
\label{geom}
\ee
Inserting this into \eqref{li} gives {\em one formula
which applies whether the target is inside or outside $\Omega$},
\be
\uap(\xi,\eta)
\;=\;
\frac{1}{2\pi}\int_{\pO} \bigl( 1 - e^{\frac{i\pi}{\lambda z}r^2} \bigr) \frac{\rr \times d\sss}{r^2}
\qquad \mbox{ (NSLI formula)~.}
\label{nsli}
\ee
This has no singularity as $r\to0$ (target approaching $\pO$)
because the term in square brackets is $\bigO(r^2)$, cancelling the denominator.
The integrand is as smooth as the Fresnel zones,
ie, as smooth as the diffracted field in the target plane.
We believe that \eqref{nsli} is new.

This leads to an incredibly simple yet robust code. For instance, in MATLAB, if
{\tt bx} and {\tt by} list coordinates of nodes on $\pO$, with {\tt wx} and {\tt wy} the
corresponding weights for a vector line integral as in Sec.~\ref{s:kite}, the entire NSLI code to output $\uap$ at a target {\tt (xi,eta)} is
five lines:

{\footnotesize
\begin{verbatim}
  rx = bx - xi; ry = by - eta;           % components of r displacement vector
  r2 = rx.*rx + ry.*ry;                  % r^2
  f = (1 - exp((1i*pi/lambdaz)*r2)) ./ r2;
  f(r2==0.0) = 0.0;                      % kill NaNs (target hits node)
  uap = sum((rx.*wy - ry.*wx) .* f) / (2*pi);      % cross product, quadrature
\end{verbatim}
}

\ch{ Note that when a target hits a node to within machine error ($r=0$),
{\em any} finite value of {\tt f} may be inserted in line 4, because
$\rr \times d\sss = 0$ in line 5.
Yet}
there is a subtlety here: numerical eyebrows should immediately be
raised because {\tt f} involves {\em catastrophic cancellation} as $r\to 0$.
\ch{ To understand why this is in fact barely a problem, we
apply forward error analysis \cite[Ch.~1]{highambook}, and}
treat the real and imaginary parts separately (combining them leads to
a pessimistic prediction).

The imaginary part of the {\tt exp} is $\sin((\pi/\lambda z)r^2)$, which,
given a rounded value of {\tt r2},
is computed to {\em relative} accuracy $\bigO(\emach)$,
where $\emach\approx 1.1\times 10^{-16}$ is the usual
double precision relative error.
Subtraction from {\tt 1} does not change the imaginary part.
The division by {\tt r2} then results in
absolute error $\bigO(\emach)$, which then gets multiplied by the
$\bigO(r)$ cross product, giving $\bigO(\emach r)$.
Note that this holds even though {\tt r2} is necessarily inaccurate
due to coordinate subtraction in line 1.

Now to the real part of the {\tt exp}, which is $1 + \bigO(r^4)$.
Thus when $r \lesssim \emach^{1/4}$, the real part of {\tt exp}
is in machine arithmetic {\em exactly} 1, which cancels the
other {\tt 1} exactly, leaving zero. Since the true answer is
$\bigO(r^3)$, in this regime the final
error is bounded by $\bigO(\emach^{3/4})$.
On the other hand, for $r \gtrsim \emach^{1/4}$, catastrophic cancellation
occurs: the error in the real part of {\tt exp} is $\bigO(\emach)$,
so the final error is $\bigO(\emach/r)$.
In summary,
uniformly in $r$, the final error is bounded by $\bigO(\emach^{3/4})$.
In practice, we find by comparison to the areal quadrature answers that this
uniform bound is around $10^{-14}$, which is adequate.
Replacing by a Taylor expansion for small $r$
\ch{ (e.g. via {\tt cexprl} \cite{cexprl}) }
could possibly gain a digit.

The formula \eqref{nsli}, in the form of the above code
looped over target points,
serves as our reference direct method.
A documented, tested MATLAB/Octave implementation is in the
repository \cite{fresnaqgit} in {\tt bdrymeths/nsli\_pts.m}

\begin{rmk}
  When a poor quadrature (that is, low order and few nodes) is used
  with deep shadow regions,
  the usual line integral \eqref{li}
  has one advantage over NSLI \eqref{nsli}:
  it can in shadows achieve {\em relative} accuracy
  in $u$, appropriate to the quadrature,
  because $u_\tbox{geom}=0$ exactly.
  NSLI merely achieves {\em absolute} accuracy in $u$, thus may
  require a better quadrature to resolve deep shadows than \eqref{li}
  (as implemented by, eg, BDWF).
  In essence, $\uapg\approx 1$ (the ``1'' term in \eqref{nsli})
  to limited accuracy, which is then poorly canceled in \eqref{bab}.
  To remedy this, our NSLI implementation also
  includes an \ch{option} to use \eqref{li} for
  targets far from $\pO$, combining the robustness of \eqref{nsli} with the
  deep shadow relative accuracy of \ch{traditional edge integrals}.
  \label{r:shadfix}
\end{rmk}

\bibliographystyle{spiejour}
\bibliography{alex}
\end{document}